\documentclass[preprint]{aastex}
\begin{document}			% REQUIRED

\title{Designing Surveys for Tests of Gravity}

\author{Bhuvnesh Jain}
\author{Department of Physics \& Astronomy, University of Pennsylvania,
 Philadelphia, PA 19104}

%\date{\today}		 		% your own text, a date, or \today
% --------------------- end of the preamble ---------------------------
%\bibliographystyle{apj}

\begin{abstract}
Modified gravity theories may provide an alternative to dark
energy to explain cosmic acceleration. We argue that the observational
program developed to test dark energy needs to be augmented to
capture new tests of gravity on astrophysical scales.  
Several distinct signatures of gravity theories exist outside the ``linear'' regime, 
especially owing to the screening mechanism that operates inside halos 
like the Milky Way to ensure that gravity tests in the solar system are satisfied.  
This opens up several decades in length scale and classes of galaxies 
at low-redshift that can be exploited by surveys.   
While theoretical work on models of gravity is in the early stages, 
we can already identify new regimes which cosmological surveys could target  
to test gravity. These include: 
1. A small scale component that
focuses on the interior and vicinity of galaxy and cluster halos.  
2. Spectroscopy of low redshift galaxies,
especially  galaxies smaller than the Milky Way, in environments that
range from voids to clusters.   3. A program
of combining lensing and dynamical information, from imaging and spectroscopic
surveys respectively, on the same (or statistically identical) 
sample of  galaxies. 
%We give a brief and qualitative discussion aimed at 
%stimulating new observational efforts to test gravity. 
\end{abstract}

\section{Introduction}

The observed acceleration in the expansion of the universe can arise from a 
dark energy component or from a departure of gravity from general relativity (GR)
on cosmological scales. 
Dark energy tests are usually phrased in terms of measuring a set of
parameters that describe the energy density and equation of state of
dark energy.  For smooth dark energy models, these are typically 
$\Omega_{\rm de}, w_0, w_a$. These parameters are independent of scale
and redshift; they describe both the expansion history, which determines
the distance-redshift relation $d(z)$ observed by Supernovae and
Baryonic Acoustic Oscillations (BAO) features in galaxy surveys, 
and the growth of perturbations.  

For modified gravity (MG) theories, there is no analog to this simple
 framework. The relation of the expansion history to the growth of
 perturbations is specific to every model\footnote{Indeed 
 scalar-tensor models typically have the counterintuitive
 combination of faster expansion of the universe (gravity acting ``weaker" 
 for the homogeneous universe) and enhanced growth of
 perturbations.}. For the quasi-static, 
 Newtonian linear regime (henceforth simply referred to as the linear
 regime), two functions of scale and redshift can describe the growth
 of perturbations for  essentially all scalar-tensor models:
\begin{equation}
g\equiv \frac{G}{G_N}, \ \ \ \ \eta\equiv \frac{\psi}{\phi} \ . 
\end{equation} 
where $G$ is the effective function that replaces Newton's constant $G_N$,  
and $\psi$ and $\phi$ are the two metric potentials. Several authors have used such a parameterization to show that
 the growth of perturbations provides useful tests of MG 
models that are  able to match the observed expansion history of the universe (see
Jain \& Khoury 2010 for a review). However in
planning cosmological surveys,  attention has mostly been limited
to the linear regime. Here we discuss potentially observable  tests that lie outside the linear 
regime -- see Figure 1 for a rough representation of various observables. 
There are two fundamental reasons for these new opportunities in 
tests of gravity:  
\begin{itemize}
\item MG theories generically break the equivalence between the mass
  distribution inferred from the motions of stars and galaxies
  (non-relativistic tracers) versus that inferred from
  photons. Thus   the comparison of dynamical and lensing masses of
  galaxies, clusters and large-scale structures can yield signatures
  of MG.  
\item MG theories rely on screening mechanisms that shield high
  density regions like the halo of the Milky Way from modified
  forces. The screening must ensure that stringent solar system and
  lab tests of GR are satisfied. Hence there is a transition
  from gravity being GR-like within large halos to 
  being modified on large scales and/or for smaller halos.   
  Potentially observable deviations  in the 
  dynamics of stars and galaxies arise from the enhanced forces in this regime. 
\end{itemize}

\section{New Regimes for Tests of Gravity}

\subsection{Comparison of Lensing and Dynamics}

The deflection law for photons that leads to various gravitational lensing effects is the same
in any metric theory of gravity. Moreover the relation of light deflection to the mass distribution
is largely unaltered in scalar-tensor theories; lensing masses are true masses. However 
the acceleration of galaxies, which move at non-relativistic speeds, is altered as the Newtonian 
potential is different from that in GR: it receives additional contributions from the scalar field in 
scalar-tensor models. Thus a discrepancy arises in the mass distribution inferred from
dynamical tracers and lensing. 

A comparison of lensing and dynamical cross-power spectra was proposed by
Zhang et al (2007) as a model independent test of gravity. This test is
in principle immune to galaxy bias, at least to first order, and can also 
overcome the limitation of sample variance on large scales. Thus it can
be applied in the linear regime 
relatively easily, provided both multi-color imaging (for lensing) and
spectroscopy (for dynamics) are available for the same sample of galaxies. Reyes et
al (2010) carried out this test with the Sloan Digital Sky Survey and found
consistency with GR.  On much smaller scales, versions of the test
can be performed by comparing  lensing and dynamical masses of halos. 
This test is fairly unique to testing gravity, as it has
little information to add in the dark energy framework. The observational
implications are discussed below.  

\subsection{A Suite of Tests in the Small Scale Screening Regime}

Two screening mechanisms 
are fairly generic to MG models: chameleon screening which is relied on by 
a variety of scalar-tensor theories (including
all $f(R)$ models), and Vainshtein screening which operates for higher 
dimensional and Galileon theories. They have been studied extensively in the literature
(Vainshtein 1972; Khoury \& Weltman 2004; see  Jain \& Khoury 2010 for a review); recent work has included detailed simulations which are necessary because of the nonlinearity
inherent in how GR is recovered inside the Milky Way. Additional screening mechanisms 
exist, such as the symmetron mechanism, but it is likely that the qualitative features in 
the small-scale regime are captured by a handful of screening
mechanisms\footnote{The effects discussed below vary between screening
  mechanisms; they  are most closely motivated by chameleon
  screening. For Vainshtein screening in the DGP model the deviations inside 
  halos are suppressed, but this expectation may change for other models or in 
  environments that do not follow the predictions for spherical halos.}. 

Screening mechanisms utilizes some measure of the mass 
 of halos, such as the Newtonian potential, to recover GR well within the Milky Way. 
It leaves open the possibility that smaller halos, such as those of dwarf galaxies, or the outer parts of bigger halos can experience enhanced forces. For a given mass distribution, 
unscreened halos will then have higher internal velocities and center of mass 
velocity compared to GR. This can produce deviations of $\sim$10-100\%  from GR, with 
distinct variations between different mechanisms in the size of the effect and the 
way the transition to GR occurs. It is important to note that observable effects are larger on 
halo scales than in the linear regime.  Since MG models recover GR at high redshift for
consistency with CMB and Nucleosynthesis observations, 
effects of enhanced forces are manifested only at late times. 
Observed velocities of stars or galaxies arise from integrating the 
acceleration over their trajectory. 
The dynamical time in galaxy halos is typically much smaller than the period over which 
modified forces (in un-screened environments) have been present. 
So an observable like the virial velocity dispersion fully reflects the modified force law. 
However large-scale perturbations have been continuously 
growing since well before the MG era, so the signal is smaller.  

Screening mechanisms and how they operate in different MG models
is an area of active research. We can nevertheless summarize some general 
features relevant to astrophysical tests: 

\begin{enumerate}

\item 
Screening mechanisms yields distinct signatures in
many settings, such as:  
i) Features in the radial profile of the mass density, 
velocity dispersion or rotation velocity, 
and infall patterns around galaxies and clusters.
ii) Variations in the dynamical masses of low mass halos (smaller than the
Milky Way) in environments with differing ambient density. 
iii) Variations in the center of mass velocities of tracers that have different masses or densities. 
These tracers can range from stars 
and gas well within galaxies to globular clusters or satellite galaxies in the
outer parts of galaxy or group halos (see Hui, Nicolis \& Stubbs 2009 for an extended discussion 
of some of these issues). 
Predictions for such tests are challenging since visible properties of galaxies may
also be affected, but  there is clear theoretical motivation to expect deviations from GR.

\item The comparisons of dynamics to lensing needs to be carried out  
differently than in the linear regime but also offers powerful tests 
on small scales. While the cross-correlation 
test of Zhang et al (2007) is targeted at large enough scales (so that
the treatment of galaxy  bias and redshift space power spectra is valid), 
 one can also directly compare
lensing and dynamical tracers of halos of galaxies and clusters (e.g. Jain
\& Zhang 2008; Schmidt 2010). Three kinds of tests are available: 
the comparison of strong lensing with measured stellar
velocity dispersions in the inner parts of elliptical galaxies (Bolton et al 2006), 
the virial masses of halos from weak lensing and dynamics, 
and the infall region that extends to ten or more times the virial radius. 
The latter two tests are feasible only for massive clusters or using stacked 
measurements of large samples of galaxies binned in luminosity or another observable 
that serves as a proxy for halo mass. 
The signal can be significantly larger than the linear regime version of the test. 

\item The screening regime yields arguably more generic tests than the
  linear or 
quasilinear regime, in the following sense. Various models of MG
that may be proposed in the future are likely to rely on one among a
handful of screening mechanisms. Therefore if an observational test
yields a signature of a particular mechanism, we have already learned
something important about what kind of MG is at
play (conversely it is less likely that a linear regime
detection of modified gravity will help identify a class of models). This reasoning 
runs counter to the conventional view that the linear regime provides the cleanest  
cosmological tests; we note that it is also somewhat speculative in that it anticipates features 
of future theoretical models. 

\item Various astrophysical uncertainties are more important 
on small scales: the relation of galaxies to halos, gas physics, velocity bias, tidal stripping and dynamical friction, and so on. It is unlikely that 
inferences about halo masses or dynamics can be obtained to better than 10\% accuracy. So one must rely on signatures that are larger than this level (which is likely to be the case if current work
is any guide). And if there are distinct 
transitions in the signal with halo radius, mass and environment, one can potentially 
extract it even in the presence of systematics that are not expected to show the same transitions. 

\end{enumerate}

{\bf Laboratory and solar system tests.} 
While local tests are beyond the scope
of this discussion (see Will 2005), it is worth noting
that astrophysical tests need to be thought of in a broader framework
that includes lab and solar system tests. As the literature on Vaishtein
and chameleon screening makes clear, the sharp constraints we already
have from local tests immediately restrict the parameter space of
models. Thus if the screening mechanisms are to be taken seriously, tests at 
different scales are not independent -- some 
set of astrophysical tests may not be interesting at the outset while others may be more 
powerful tests of particular screening mechanisms than local tests, even though their absolute accuracy is much lower.

\section{Large Scale Tests: Limitations}

%\subsection{The Linear Regime of Perturbations} 

As with tests of dark energy, the linear regime offers 
advantages in the ease of predictions and interpretation, 
and immunity from astrophysical systematics which typically cannot 
alter structure formation on scales above 100 Mpc. And MG models can 
produce scale and redshift dependent growth that distinguishes them 
from dark energy models with the same expansion history. 
However there are some limitations which are exacerbated for tests of
gravity\footnote{The ISW effect offers useful tests of gravity that
  may be exceptions to the linear regime discussion here. However ISW
  measurements are not likely to improve much with upcoming surveys due 
  to cosmic variance limitations.} : 

\begin{enumerate}

\item Upcoming surveys with Stage III capabilities (in the terminology of the Dark Energy Task Force) will be limited by sample variance to 
$\Delta P/P \sim 10\%$ level  errors for power spectra of interest such as the galaxy-velocity
cross-spectrum.   Systematic errors of various kinds 
are difficult to control: it remains to be seen if their  contribution to $\Delta P$ is below the MG signal on the largest scales. 

\item The regime has limited range in length scale. 
The $g, \eta$ parameterization of Eqn. 1 is 
useful only on scales sufficiently smaller than the superhorizon regime and
larger than the nonlinear/screening regime. 
This demarcation is model dependent. 

\item At high-redshift the linear regime spans a wider range of scales,
  but the signal is also smaller. In this respect, 
  the growth of perturbations differs
  from the distance-redshift relation: the signal accumulates from high
  redshift (where MG is suppressed) rather than $z=0$.  Moreover 
the redshift range at which MG effects kick in can depend on scale -- 
for $f(R)$ models it is smaller for larger scales, restricting the useful linear regime further. 

\end{enumerate}

%While the first point above applies to dark energy tests as well, 
%the second and third apply more sharply to gravity tests. It may well
%be that linear regime tests with strong discriminating capability 
%will need to wait another decade for Stage IV survey data. 
%Thus the extension of a PPN-like framework for linear regime tests of gravity has some %limitations: the parameters are not constant with either redshift or scale, and their behavior is well %understood over a limited dynamic range. 

{\bf Weakly Nonlinear Regime:} 
The few-100 Mpc regime is observationally easier because of higher signal-to-noise, 
but requires more work to model. 
Accurate computations of nonlinear gravitational clustering require 
N-body simulations which are specific to every model of MG. There is some 
hope that using a $g,\eta$ approach 
a suite of simulations can be used to calibrate analytical formulae the same way as GR, 
just with a modified linear growth factor.  However, baryonic physics, galaxy bias and
nonlinear screening effects which operate at a low level need to be
modeled or marginalized with care. While nonlinear screening effects 
are not expected to play a significant role on scales above 10 Mpc, this has to be 
established for each model and observable (e.g. in Fourier space it is not straightforward 
to find the corresponding wavenumber). 
The size of MG effects may be comparable to the
linear regime (i.e. small) for some observables, but  larger for others such as the  
halo mass function in $f(R)$ models. Thus the quasilinear regime offers 
opportunities for tests of gravity even with Stage III surveys, though it is probably 
easier to regard them as null tests of GR rather than finding signatures of particular 
MG models. 

\section{Implications for Cosmological Surveys}

\begin{enumerate}

\item As discussed in Section 2, 
modest sized galaxies at small separations can provide
useful tests of screening mechanisms.  For galaxies smaller than the Milky Way, 
it is especially useful to compare their dynamics in environments ranging from
voids to galaxy groups and clusters. Since 
spectroscopic surveys typically optimize their resources to get
galaxy pairs in the 10-100 Mpc range for BAO tests, tests of gravity 
may require a different observational strategy. 
%For the higher mass halos it would be useful to study dynamics at
%different radii. 

\item The low redshift universe is likely to provide stronger tests of gravity, 
at least for Stage III surveys. Whereas dark energy tests
   using BAOs benefit from $z\sim 1$ data because additional peaks at high-$k$
   fall into the linear regime, for modified gravity the signatures
   with redshift are harder to detect observationally than the wealth
   of other information at low-$z$. (Well designed Stage IV surveys
   may well cover the entire redshift range of interest for dark
   energy/MG tests: they are likely to have the depth and area/volume
   to carry out linear regime tests at high-$z$,  as well as many low-$z$
   tests.) 

{\it A consequence of the above two points is that pursuing dwarf 
galaxies at low-$z$ is arguably as useful for gravity tests as intrinsically
brighter galaxies at high-$z$ (with 
a comparable limiting magnitude). 
Since low-z dwarf galaxies are not a part of the
strategy for spectroscopic surveys aiming for BAO measurements, the 
resources directed at high-$z$ galaxies 
may need to be partially re-allocated for tests of gravity. 
Note that the spatial sampling of the low-z galaxies need not be  uniform over the 
sky, instead it is probably useful to obtain large subsamples of  galaxies
in different environments.  Such observations can be carried out as part of 
cosmological surveys, and also by spectroscopic cameras with modest fields of view 
that are unable to cover the wide areas needed for dark energy
surveys. Several parameters of the galaxy population, instrument and survey 
strategy need to be studied to implement useful tests. 
Imaging surveys require less modification for gravity tests since they 
typically cover large contiguous areas and a wide range
in redshift; even so, low-z groups and clusters may need to be observed under 
different conditions or with wider filter coverage than planned. }  

\item Tying together lensing and dynamical information from 
imaging and spectroscopy requires identical, or statistically indistinguishable,  
samples of lens/host galaxies. It remains to be established how hard it is to 
get two samples of statistically identical galaxies
 if the two surveys don't overlap. Can
calibration issues (filters, extinction, seeing) 
undermine even the most scrupulous 
color and magnitude selection? 
%This issue may become problematic if 
%current trends of imaging surveys in the South and spectroscopy in the
%North continue! 

\item In addition to wide area imaging and spectroscopic surveys,
   supplementary observations are likely to be useful. For instance high
   resolution spectroscopy of strong lensing galaxies can provide useful measures of 
   stellar velocity dispersion. Tests based on the properties of populations of stars may also be 
   useful since modified forces can alter stellar evolution 
   (Chang \& Hui 2010; Davis et al 2011). We have been 
concerned with cosmological surveys here, so we have not considered these effects. 
However they may be relevant in interpreting some of the dynamical data 
discussed above: altered stellar evolution could impact the mass-to-light properties 
of galaxies and the fundamental plane or Tully-Fisher relations. 
And we have only touched on solar system 
  and laboratory tests above, but there is obviously a close coupling of information 
  from these tests and astrophysical ones. 
  
\end{enumerate}

\section{Discussion and Caveats}

The brief  discussion above has been aimed at stimulating new 
observational approaches to test gravity. But 
currently there are no ``successful'' MG models, so how
confident can one be about the recommendations made here? 
As far as possible we have relied  on generic features of screening
mechanisms that operate for classes of models, but 
it is fair to say that theorists are in the early stages of building MG models 
and working out observational consequences. 
More work is needed to determine how model-independent various tests of
screening mechanisms really are.  

While there are caveats to any predictions  made by MG theories, 
there are some general reasons to expect that such theories require 
a scalar field that leads to force enhancements on observable scales 
(the argument is along the lines of Weinberg 1965). 
Current work suggests the enhancement is larger than 10\% and  
possibly approaches  the 100\% level outside of screened halos. 
It is difficult to be specific about the best range of galaxy
luminosity or radii 
to probe the enhanced forces. But simply by using the Milky Way as a
reference point, one can be confident that  pushing 
observations to smaller galaxies, comparing results in a diversity of
environments, and combining lensing and dynamics wherever possible
will yield interesting tests. More work is needed to formulate specific tests  
and the best strategy for obtaining useful galaxy samples.

Alternatively, one may disregard recent theory and simply continue the decades-long program of 
testing GR, but then astrophysical tests are not well motivated: deviations from GR 
in the Parameterized Post-Newtonian approach are independent of scale, so local tests 
measure parameters of interest with much more precision. 
The recent theoretical work discussed above  
has shown that the potentially large dependence of deviations from GR 
on scale and environment  makes astrophysical tests as discriminating as local tests. 
Theoretical input is thus valuable in suggesting what regimes to  
test and what precision is likely to be useful -- it informs choices that have to be made in
allocating observational resources.   

Some of the discussion of MG models may also be taken to apply to interactions in the
dark sector: scalar-tensor gravity theories can be rephrased  as GR 
containing a scalar field with couplings to matter. Even in a dark energy scenario, 
one may argue that it is not unreasonable to expect a dynamical dark energy to couple to  
matter. The field associated with dark energy would then need to rely on 
chameleon screening to satisfy
equivalence principle tests on earth. Tests of the  screening regime then become
tests of such couplings. So as long as $\Lambda$CDM with GR is not considered a 
compelling description of our universe, these tests remain of interest.

Finally, in planning surveys one must  recognize that MG
scenarios are still being developed, and are complex in the interesting
sense of having a diversity of predictions. There is no 
simple figure of merit to describe tests of gravity, and there are
good reasons to expect that this will remain the case. 
Not only is there a huge range in length scale outside
of the linear regime, but properties of the tracer
and its environment can be important. So one 
must address several different criteria with their own metrics rather than a single
Figure of Merit. One can hope that  
future work that incorporates new ideas on MG models 
will make these criteria more concrete.

{\it Acknowledgements:} Much of what I know about screening mechanisms
I have learned from Justin Khoury. I am grateful to him for sharing
his insights and for joint work on a review article on gravity. I
acknowledge stimulating discussions and feedback from Gary Bernstein, 
Anna Cabre, Joseph Clampitt, Jacek Guzik, Kurt Hinterbichler, Lam Hui, 
Mike Jarvis, Eric Linder, Fabian Schmidt, Roman Scoccimarro,
Ravi Sheth, Masahiro Takada and Mark Trodden. 

\begin{figure}
\includegraphics[width=12cm]{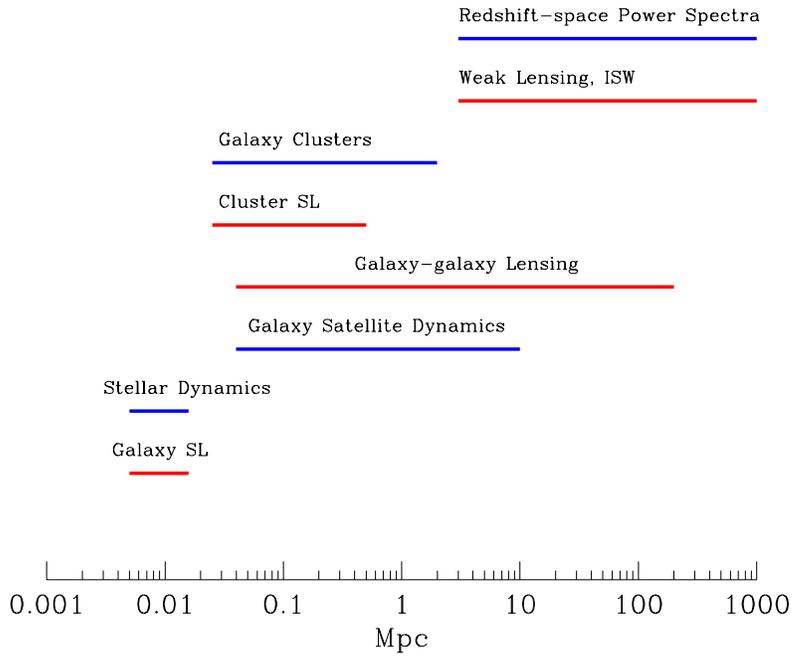}
\caption{Tests of gravity at different length scales (adapted from Jain \& Khoury 2010). 
Red lines shows observations
that probe the sum of metric potentials via weak and strong gravitational lensing (SL) or
the ISW effect. Blue lines show dynamical measurements that rely on the motions
of stars or galaxies or other non-relativistic tracers. This partial list of observables 
illustrates the wide range of scales that can provide interesting tests. In addition, properties
of the tracer and its environment are also important. 
}
\label{fig:GravityScales}
\end{figure}

\section*{References}
B. Jain and J. Khoury, Annals of Physics, {\bf 325}, 1479 (2010)

P. Zhang, M. Liguori, R. Bean, and S. Dodelson, 
 Phys.Rev.Lett.  {\bf 99}, 141302 (2007)  

R. Reyes et. al. %, R., Mandelbaum, 
%R., Seljak, U., Baldauf, T., Gunn, J.~E., Lombriser, L., 
%\& Smith, R.~E.\ 2010, 
Nature, 464, 256 (2010)

A.~I.~Vainshtein,
% ``To the problem of nonvanishing gravitation mass,''
Phys.\ Lett.\ B {\bf 39}, 393 (1972).

  J.~Khoury and A.~Weltman,
%  ``Chameleon fields: Awaiting surprises for tests of gravity in space,''
  Phys.\ Rev.\ Lett.\  {\bf 93}, 171104 (2004)

 L.~Hui, A.~Nicolis and C.~Stubbs,
%  ``Equivalence Principle Implications of Modified Gravity Models,''
Phys. Rev. D, 80, 4002 (2009)

B. Jain, \& P. Zhang,  Phys. Rev. D, 78, 063503 (2008)

F. Schmidt, Phys. Rev. D, 81, 103002 (2010)

A. S. Bolton, S. Rappaport  \& S. Burles , Phys. Rev. D, 74, 061501 (2006)

  C.~M.~Will,
  %``The confrontation between general relativity and experiment,''
  Living Rev.\ Rel.\  {\bf 9}, 3 (2005)
 % [arXiv:gr-qc/0510072].

P. Chang, \& L. Hui, arXiv:1011.4107 (2010)

A.-C. Davis, A. Lim, 
J. Sakstein, \& D. Shaw, arXiv:1102.5278 (2011)

S. Weinberg, Phys. Rev. B, 4, 138 (1965)

%\bibliography{lensing}
%\begin{thebibliography}

%\bibitem[Aller \& Reynolds(1985)]{aller} Aller, H.~D., \& Reynolds, S.~P.\ 1985, \apjl, 293, L73 

%\bibitem[tmp]{review}
%B. Jain and J. Khoury  2010, Annals of Physics, {\bf 325}, 1479 %(2010)

%\end{thebibliography}

\vfill 

\end{document}